\documentclass{SIP}

\usepackage{booktabs}
\usepackage{tabularx}
\newcolumntype{Y}{>{\centering\arraybackslash}X}

\def\BibTeX{{\rm B\kern-.05em{\sc i\kern-.025em b}\kern-.08em
    T\kern-.1667em\lower.7ex\hbox{E}\kern-.125emX}}
\usepackage{epstopdf, epsfig}

\begin{document}

\title [Neural Post-filter]{A Cyclical Approach to Synthetic and Natural Speech Mismatch Refinement of Neural Post-filter for Low-cost Text-to-speech System}

\author[Yi-Chiao Wu, \textit{et al}.]{Yi-Chiao Wu$^{1}$, Patrick Lumban Tobing$^{1}$, Kazuki Yasuhara$^{2}$, Noriyuki Matsunaga$^{2}$, Yamato Ohtani$^{2}$ and Tomoki Toda$^{1}$}

\address{\add{1}{Information Technology Center, Nagoya University, Furocho, Chikusa Ward, Nagoya, Aichi 464-8601 Japan}
\add{2}{AI, Inc., 
W Building 2F, 2-2-2 Hikaridai, Seika-cho, Souraku-gun, Kyoto 619-0237, Japan}}

\corres{\name{Yi-Chiao Wu}
\email{yichiao.wu@g.sp.m.is.nagoya-u.ac.jp}}

\begin{abstract}
Neural-based text-to-speech (TTS) systems achieve very high-fidelity speech generation because of the rapid neural network developments. However, the huge labeled corpus and high computation cost requirements limit the possibility of developing a high-fidelity TTS system by small companies or individuals. On the other hand, a neural vocoder, which has been widely adopted for the speech generation in neural-based TTS systems, can be trained with a relatively small unlabeled corpus. Therefore, in this paper, we explore a general framework to develop a neural post-filter (NPF) for low-cost TTS systems using neural vocoders. A cyclical approach is proposed to tackle the acoustic and temporal mismatches (AM and TM) of developing an NPF. Both objective and subjective evaluations have been conducted to demonstrate the AM and TM problems and the effectiveness of the proposed framework. 
\end{abstract}


\maketitle

\section{Introduction}
\label{sec:introduction}
Text-to-speech (TTS) is a technique to generate speech with specific contents corresponding to the input texts. Because of the developments of deep learning and neural networks, the state-of-the-art TTS systems are dominated by neural-based TTS systems~\cite{tacotron2, fastspeech, fastspeech2}. A typical neural-based TTS system is composed of acoustic and speech generation modules. The acoustic module converts input text to acoustic features such as mel-spectrogram. The speech generation module, which is known as a vocoder, generates speech waveforms on the basis of the predicted acoustic features.

Although the state-of-the-art neural-based TTS systems achieve very high fidelity speech synthesis, a huge labeled corpus and high computation costs are usually required. Because collecting high-quality labeled data and training the complicated neural networks are quite expansive, and the requirement of a huge amount of data for a new language or user is impractical, building a state-of-the-art TTS system from scratch is difficult for small companies and individuals. Therefore, many small companies tend to build a lightweight TTS system such as the hidden Markov model (HMM)/ deep neural network (DNN)-based Speech Synthesis System (HTS)~\cite{hts} using a limited amount of labeled data or directly adopt existing TTS systems. However, the limited quality of these low-cost options hinders the practicality of these systems for practical applications. 

To tackle the issue, a straightforward strategy is to develop a neural post-filter (NPF) to improve the generated speech quality of these low-cost and simple TTS systems. Specifically, only a relatively small unlabeled corpus is required for training a high-quality neural vocoder ~\cite{sd_wn_vocoder, si_wn_vocoder, ns_wn_vocoder} because the input acoustic features can be directly extracted from the output waveforms without any labels, and only the relatively simple acoustic features to waveforms mappings are performed. Therefore, training a neural vocoder as the post-filter to enhance the synthesized speech from the low-cost or existing TTS systems is economical and feasible for small businesses

However, there are two main problems, acoustic and temporal mismatches (AM and TM), for developing an NPF for arbitrary TTS systems. Specifically, if an NPF is trained using natural acoustic features and waveforms but tested using the synthetic acoustic features extracted from the generated speech of a low-cost TTS system, the different acoustical characteristics between the natural and synthetic features cause a severe AM problem. Furthermore, in our previous work~\cite{cycpf_2020}, we argue that even if an NPF is trained using the synthetic features and natural waveforms, the different temporal structures between the synthetic features and natural waveforms also cause a serious TM problem. The experimental results in~\cite{cycpf_2020} show that both AM and TM problems result in significant speech quality degradation.

To tackle the AM and TM problems, we adopted a cycle voice conversion (Cycle-VC)~\cite{cyc_vc} model to generate pseudo-VC features for NPF training and enhanced features for NPF testing. The Cycle-VC model includes two paths to respectively convert synthetic features to natural features (enhanced path) and natural features to synthetic features and back to natural features (pseudo-VC path). Since the pseudo-VC features are converted from natural features, the temporal structures of the pseudo-VC features and natural waveforms are matched (TM-free). Both the pseudo-VC and enhanced features are processed by the Cycle-VC model, so they have similar acoustical characteristics (AM-eased). 

To show the effectiveness of the cyclical approach, an autoregressive (AR)-based generative model, WaveNet~\cite{wavenet}, was adopted as the NPF model and achieved marked improvements in~\cite{cycpf_2020}. However, several remaining problems and questions should be addressed and answered. First, since not only AR vocoders but also non-AR vocoders have been widely adopted in current speech generation applications~\cite{vcc2020, blizzard2020}, it is important to explore the effectiveness of the proposed framework utilizing a non-AR vocoder. Second, only the effectiveness of the proposed method for testing utterances with oracle phoneme durations has been explored, but only predicted phoneme durations are available in the testing stage. Third, the widely adopted conventional spectral post-filter~\cite{hts, vc_mlpg, hmm_pf} included in HTS is very effective, but the further comparisons between the proposed method and the conventional post-filter have not been explored  

To address these problems, we extend our previous work in this paper. There are four new contributions.
\begin{itemize}
\item We further investigate the effectiveness of the proposed cyclical approach to a non-AR generative model, parallel WaveGAN (PWG)~\cite{pwg}, and the experimental results demonstrate the versatility of the cyclical approach for both AR and non-AR generative models. The different detailed tendencies of WN and PWG vocoders in the proposed framework also give us more insight into the AR and non-AR vocoders.
\item We further explore the effectiveness of the proposed NPF for the real scenario that the oracle durations are not adopted for the testing synthetic speech. Two training strategies, using oracle or predicted phoneme durations in the training stage, are also explored to provide the training paradigm to build a practical NPF.
\item We further compare the proposed NPF with the conventional spectral post-filter included in HTS, and the proposed NPF significantly outperforms the conventional post-filter according to the experimental results. Furthermore, the effectiveness of the proposed NPF combined with the conventional post-filter is also evaluated. 
\item We further provide visualized and objective experimental results to demonstrate the AM and TM problems since the new concept of the TM problem is not very straightforward when the oracle phoneme durations have been utilized in the testing stage.
\end{itemize}

This paper is organized as follows. In Section~\ref{sec:related}, we review the recent neural vocoders, the post-filter techniques for TTS, and the acoustic mismatch refinement methods. In Section~\ref{sec:vocoder}, a brief introduction to AR-based WN and non-AR-based PWG are presented. In Section~\ref{sec:cyc}, we describe the concepts and details of the Cycle-VC. In Section~\ref{sec:npf}, the AM and TM problems are explained, and the proposed cyclical approach for NPF is described. In Section~\ref{sec:experiment1}, we report objective and subjective experimental results to demonstrate the AM and TM problems and evaluate the effectiveness of the proposed NPFs for the synthetic speech using oracle phoneme durations. In Section~\ref{sec:experiment2}, we further evaluate the proposed framework in a real-world scenario that the phoneme durations of the synthetic speech are also predicted. Finally, the conclusion is given in Section~\ref{sec:conclusion}.

\section{Related Work}
\label{sec:related}
\subsection{Neural speech generative model}
Neural speech generative models generate raw speech waveform samples on the basis of acoustic features such as mel-spectrogram or lingustic featurees. One of the main challenges is to recover the phase information which is usually not included in the input acoustic or linguistic features. AR models like WaveNet~\cite{wavenet} and SampleRNN~\cite{samplernn} achieve early success in speech generation because of the autoregressively sample by sample generation well handing the phase recovery. However, the AR mechanism and complicated network architectures result in a very slow generation.

To improve the generation efficiency, recurrent neural network (RNN)-based WaveRNN~\cite{wavernn} and LPCNet~\cite{lpcnet} have been proposed. Both of them adopt very simple network architecture with a few RNN layers to achieve a real-time generation by greatly reducing the computation costs. To reduce the required modeling capacity of the generative network, WaveRNN adopts mel-spectra as strong prior information with several hardware-efficient designs, and LPCNet makes the network focus on only the residual signal which is almost much simple and speaker-independent.

On the other hand, many convolutional neural network (CNN)-based models have also been proposed to achieve real-time inference by taking advantage of the parallel computation of CNN. For example, normalize flow-based models~\cite{pwn, clarinet, wavevae, waveglow, flowavenet, waveffjord} adopt invertible network architectures to bidirectionally transfer speech signal and Gaussian distributions in the training and inference stages. The direct distribution transformations implicitly handle the phase information. Generative adversarial network (GAN)-based models~\cite{pwg, melgan, gantts} adopt discriminators to implicitly guide lightweight generators to handle the phase information.

In this paper, to show the versatility of the proposed framework, we evaluate the proposed framework using the state-of-the-art AR-based WaveNet vocoder with noise shaping~\cite{sd_wn_vocoder, si_wn_vocoder, ns_wn_vocoder} and the non-AR-based lightweight PWG vocoder which have different training and inference mechanisms.

\subsection{Post-filter for TTS}
One of the main problems causing quality degradations of statistical parametric speech synthesis (SPSS) is the over-smoothing~\cite{spss_2009} of the synthetic acoustic features. To address the problem, many post-filters have been proposed to enhance the synthetic acoustic features. For example, the post-filters in~\cite{celp_1995} and~\cite{hmm_pf} focus on enhancing the spectral formats. In~\cite{gv_pf_2007} and~\cite{vc_mlpg}, the trajectories and global variances of the synthetic spectral features have been modified to match that of the natural speech. A modulation-spectrum-based method~\cite{ms_pf_2014} was also introduced in SPSS for matching the power spectrum of the synthetic acoustic features to that of the natural acoustic features.

Following the same idea of converting the synthetic acoustic features to match the natural ones, many neural-based methods also have been proposed because of the powerful modeling ability of DNNs. For instance, different neural-based models such as feedforward DNN~\cite{dnn_pf_2014, dnn_pf_2015}, RNN~\cite{rnn_pf_2016}, long short-term memory (LSTM)~\cite{lstm_pf_2018}, and GAN~\cite{gan_pf_1, gan_pf_2} have been adopted to enhance all or partial synthetic acoustic features. Furthermore, instead of post-filtering the synthetic acoustic features, the authors of~\cite{wave_cyc_gan_1} and~\cite{wave_cyc_gan_2} also proposed GAN-based post-filters to directly enhance the synthetic speech in the waveform domain.

However, most of the acoustic feature-based methods are not jointly optimized with the downstream vocoders, and the training-testing mismatches usually result in quality degradations. Moreover, because of the high temporal resolution of waveform signals, directly manipulating waveform signals like the waveform-based methods is challenging and usually results in unstable training or inference. Therefore, in this paper, we first explore two mismatch problems, acoustic and temporal mismatches, which are caused by independently training feature conversion and vocoder modules. Then, we propose a general framework including both a synthetic spectral feature enhancement and vocoder fine-tuning for enhanced spectral features.

\subsection{Mismatch refinement for downstream vocoder}
In addition to TTS, VC is another speech generative task frequently using vocoders. Because of the data length differences between source and target utterances, an additional alignment mechanism is required for fine-tuning neural vocoders using converted utterances, whose lengths are usually the same as that of the source utterances, and target utterances. To avoid alignment errors, many methods have been proposed to generate pseudo-VC utterances, which have similar acoustic characteristics to the converted utterances but the same data lengths as the target utterances, for neural vocoder fine-tuning. For example, a Gaussian mixture model (GMM)-based intra-speaker conversion model has been adopted to fine-tune a WaveNet vocoder~\cite{vc_wn_gmm}. LSTM-based~\cite{vc_wn_lstm}, variational autoencoder (VAE)-based~\cite{vc_wn_vae}, and cyclic gated recurrent unit (GRU)-based~\cite{cyc_vc} intra-speaker conversion models have also been explored for neural vocoder fine-tuning. 

In this paper, to tackle the AM and TM problems, we adopt a cycle-VC model for post-filtering the synthetic features and producing pseudo-VC data for vocoder fine-tuning.

\section{Neural Vocoder}
\label{sec:vocoder}
\begin{figure}[t]
\begin{center}
\includegraphics[width=0.85\columnwidth]{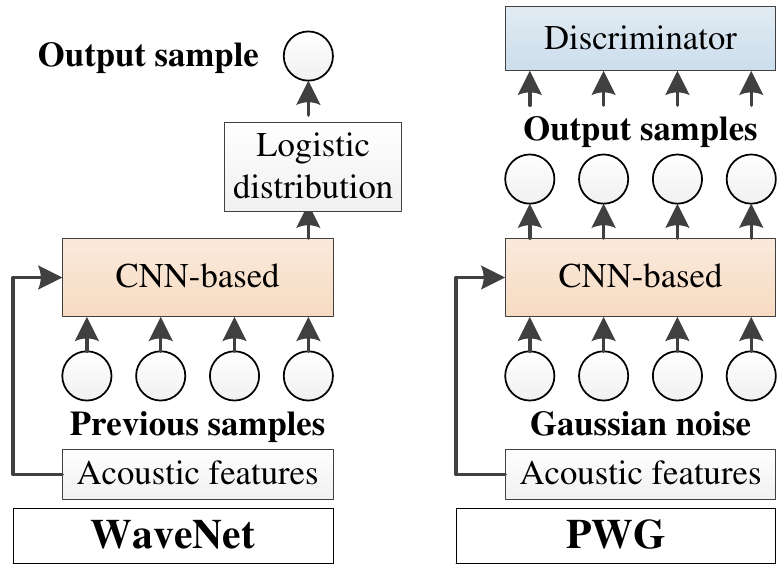}
\caption{WaveNet and Parallel WaveGAN vocoders}
\label{fig:vocoder}
\end{center}
\end{figure}
A vocoder~\cite{vocoder_1939, vocoder_1966, phase_vocoder} is a voice coder that analyzes speech into acoustic features and synthesizes speech on the basis of the acoustic features. However, a neural vocoder~\cite{sd_wn_vocoder, si_wn_vocoder, ns_wn_vocoder} typically refers to only a neural-based speech generative model synthesizing speech waveforms given input acoustic features. In this paper, two CNN-based neural speech generative models, WN~\cite{wavenet} and PWG~\cite{pwg}, are adopted as the neural vocoders for the proposed NPFs. The details are following.

\subsection{WaveNet}
WN is a fully convolutional network that directly models the conditional probability distribution of each speech sample with an AR manner. Specifically, because speech is a high-temporal resolution sequential signal with very long-term dependencies, the main concept of WN is that the current sample is related to a specific number of previous samples, which is called a receptive field, and the content of the speech sequence can be controlled using auxiliary features. Given a waveform sequence $ \boldsymbol{x}$ and an auxiliary feature sequence $ \boldsymbol{h}$, the output of WN is a probability distribution formulated as
\begin{align}
P\left ( \boldsymbol{x}\mid\boldsymbol{h} \right )
=\prod_{t=1}^{T}P\left ( x_t\mid x_{t-1},\ldots,x_{t-r},\boldsymbol{h} \right ),
\label{eq:c_prob}
\end{align}
where $t$ denotes the sample index and $r$ is the length of the receptive filed.

As shown in Fig.~\ref {fig:vocoder}, the inputs of a WN model are the previous samples, and the output is the logistic distribution of the current sample. To limit the possible values of each sample for a tractable magnitude of the logistic distribution, 8-bit $\mu$-law is applied to speech samples. Acoustic features are adopted as the auxiliary features to control the generated contexts. The CNN-based network architecture consists of several residual blocks, and each residual block includes a dilated CNN layer~\cite{dcnn}, a residual connection~\cite{residualcon}, a skip connection, a gated structure~\cite{gated}, and conditional auxiliary features. The summation of all skip connections is processed by several CNN layers with ReLU~\cite{relu} activations and a final softmax layer to output the final logistic distribution. The WN training process can be efficiently done using the teacher forcing~\cite{teacherforcing} algorithm. However, the generation follows the AR process (sample-by-sample generation), which usually results in a slow generation.

\subsection{Parallel WaveGAN}
As shown in Fig.~\ref {fig:vocoder}, PWG is a standard GAN~\cite{gan} architecture that consists of a CNN-based speech generator and a CNN-based discriminator. The generator adopts a WN-like structure without the AR mechanism to parallelly convert an input Gaussian noise sequence into a speech waveform sequence conditioned on the auxiliary acoustic features. Because of the non-AR mechanism, PWG achieves an efficient generation faster than real-time. The discriminator is also composed of several dilated CNN layers but without the dilation repeat in WN. The activations in the discriminator are LeakyReLU~\cite{leakyrelu}. Least-squares GAN~\cite{lgan} is adopted for PWG because of its training stability. To further improve the training stability and efficiency, a multi-resolution short-time Fourier transform (STFT) auxiliary loss~\cite{pwg} is also applied to the PWG training.

\section{Cyclical Spectral Conversion}
\label{sec:cyc}

\begin{figure}[t]
\begin{center}
\includegraphics[width=0.85\columnwidth]{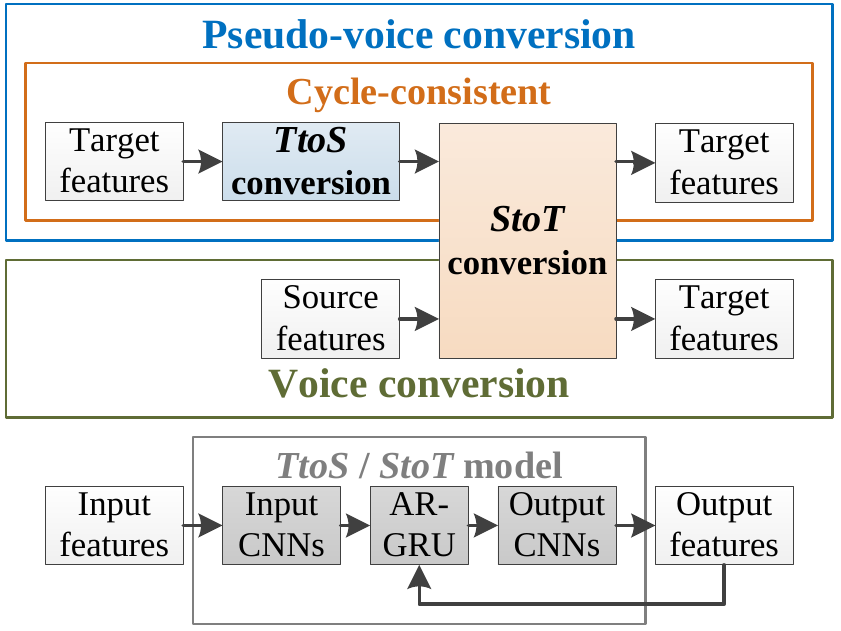}
\caption{Cycle-VC system}
\label{fig:cycvc}
\end{center}
\end{figure}

As shown in Fig.~\ref{fig:cycvc}, a Cycle-VC model includes two conversion paths, VC and pseudo-VC. The main motivations of the Cycle-VC model are 1) improving the VC performance by using cycle-consistent and 2) generating alignment-free converted features for fine-tuning the downstream neural vocoder. Specifically, a standard VC model typically includes only the source to target ({\it StoT}) spectral conversion module to convert source features to target features, and the final converted speech is generated using an independent neural vocoder, which is usually separately trained using natural feature-waveform pairs. Although the training and testing mismatch (AM of the input features) causes quality degradations, directly training the vocoder using converted features requires an additional alignment between the converted features and natural waveforms, which usually results in more severe quality degradations. 

To tackle these problems, an additional target to source ({\it TtoS}) spectral conversion module is applied to the pseudo-VC path of the Cycle-VC model. Specifically, given a source spectral vector $\boldsymbol{A}=\left [ \boldsymbol{a}_{1}^{\top},\cdots, \boldsymbol{a}_{n}^{\top} \right  ]^{\top}$, a target spectral vector $\boldsymbol{B}=\left [ \boldsymbol{b}_{1}^{\top},\cdots, \boldsymbol{b}_{n}^{\top} \right  ]^{\top}$, a nonlinear function $ f_{st}$ from the {\it StoT} module, and another nonlinear function $f_{ts}$ from the {\it TtoS} module, the loss function of the Cycle-VC model is formulated as
\begin{align}
\mathop{\arg\min}_{\theta, \phi}(\|f_{st}(\boldsymbol{A})-\boldsymbol{B}\|_{L1} + \rho\|f_{st}(f_{ts}(\boldsymbol{B}))-\boldsymbol{B}\|_{L1}),
\label{eq:cycvc}
\end{align}
where $\theta$ and $\phi$ are respectively the model parameters of the {\it StoT} and {\it TtoS} modules. $\left \| \cdot  \right \|_{L1}$ is the L1 norm. $\rho$ is a hyper-parameter to prevent the network from being dominated by the self-conversion. In this paper, $\rho$ is empirically set to $1e^{-8}$.

Both the {\it StoT} and {\it TtoS} modules are frame-based spectral conversion models composed of input CNN layers, AR-GRU blocks, and output CNN layers as shown in Fig.~\ref{fig:cycvc}. The input CNN layers consist of a {$1\times1$} CNN and two {$3\times1$} CNNs with a dilation size of 3. The AR-GRU includes one hidden layer with 1024 hidden units. The output CNN layers consist of two {$1\times1$} CNNs. In this paper, the same {\it StoT} and {\it TtoS} modules from our previous work~\cite{cyc_vc} are adopted for the spectral conversions.

According to the experimental results in~\cite{cyc_vc}, the cycle-consistent constraint provided by the pseudo-VC path improves the VC capability of the {\it StoT} module. Fine-tuning the vocoder using the alignment-free pseudo-VC features and natural waveforms greatly eases the AM problem.

\section{Proposed Cyclical Approach to Developing Neural Post-filter for TTS}
\label{sec:npf}
In this section, the AM and TM problems are first introduced, and then the proposed cyclical framework for training and testing an NPF for arbitrary TTS systems is presented. The details are as follows.

\subsection{Acoustic and temporal mismatches}

\begin{figure}[t]
\begin{center}
\includegraphics[width=0.75\columnwidth]{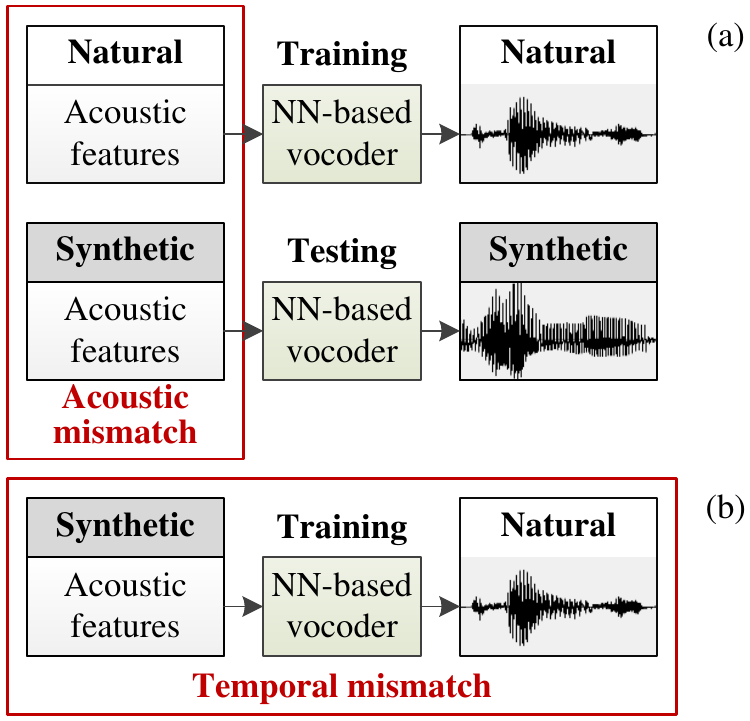}
\caption{Acoustic and temporal mismatches}
\label{fig:am_tm}
\end{center}
\end{figure}

Since speech is a sequential signal with a very high temporal resolution and long-term dependencies, directly modeling and manipulating speech in the time domain is challenging. Therefore, many speech generation applications manipulate speech in a low-dimensional feature space such as mel-spectrogram and then generate the final speech waveform using a neural vocoder. In this paper, the proposed post-filter framework follows the same design paradigm to extract acoustic features from the TTS-generated synthetic speech and then generate the final enhanced speech on the basis of the modified acoustic features using a neural vocoder.

However, there are two problems, AM and TM, as shown in Fig.~\ref{fig:am_tm}. First, if the neural vocoder is trained using natural feature-waveform pairs, the AM between the training natural and testing synthetic features usually causes significant quality degradations. That is, the synthetic features are unseen in the vocoder training stage, and their acoustic characteristics are very different from that of the natural features. Since data-driven neural vocoders are usually vulnerable to unseen testing data, the generated speech affected by the AM problem usually suffers from the oversmoothing, noisy, and unexpected sound problems~\cite{phdthesis}.

Secondly, even if the neural vocoder is directly trained using the pairs of the synthetic features and natural waveform, the TM between the synthetic features and natural waveform still causes serious quality degradations. Specifically, for learning-based TTS systems, since the duration of each speech unit such as phoneme is predicted by the model, the synthetic and natural speech waveforms are not entirely aligned resulting in TMs. However, we argue that even if the oracle durations are given, the TTS-generated speech still has some different temporal structures from the natural speech. That is, the aligned accuracy of the synthetic and natural waveforms depends on the length of the speech unit defined by the TTS system, so some detailed temporal differences may exist within each speech unit. Training a neural vocoder using pairs suffering the TM problem usually degrade the mapping accuracy of the neural vocoder resulting in unstable and unexpected sounds~\cite{cycpf_2020}. More details of the TM problem will be presented in Section~\ref{sec:exp_obj}.

\subsection{Cyclical approach to mismatch refinement}

To tackle the TM problem, the temporal structures of the input acoustic features should be consistent with that of the natural acoustic features. To tackle the AM problem, the acoustic characteristics of the training and testing input acoustic features should be similar. In this paper, we propose a cyclical model to respectively convert the natural features to pseudo-VC features for training the NPF and the synthetic features to enhanced features for testing the NPF. Since the pseudo-VC features are converted from the natural features, the temporal structures of the pseudo-VC features match that of the natural waveform. Since both the pseudo-VC and enhanced features are converted by the cyclical model, their acoustical similarity is more than that between the natural and synthetic features.

\begin{figure}[t]
\begin{center}
\includegraphics[width=1.0\columnwidth]{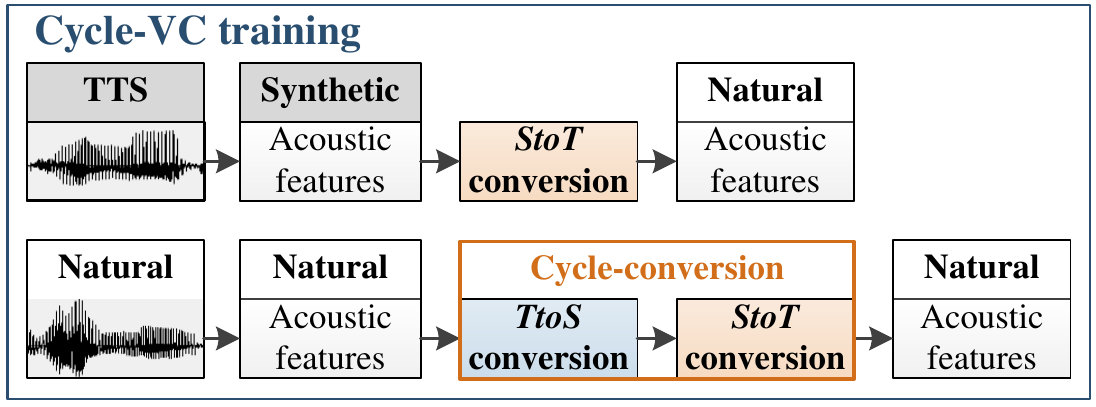}
\caption{Cycle-VC training stage}
\label{fig:cycvc_train}
\end{center}
\end{figure}

\begin{figure}[t]
\begin{center}
\includegraphics[width=0.95\columnwidth]{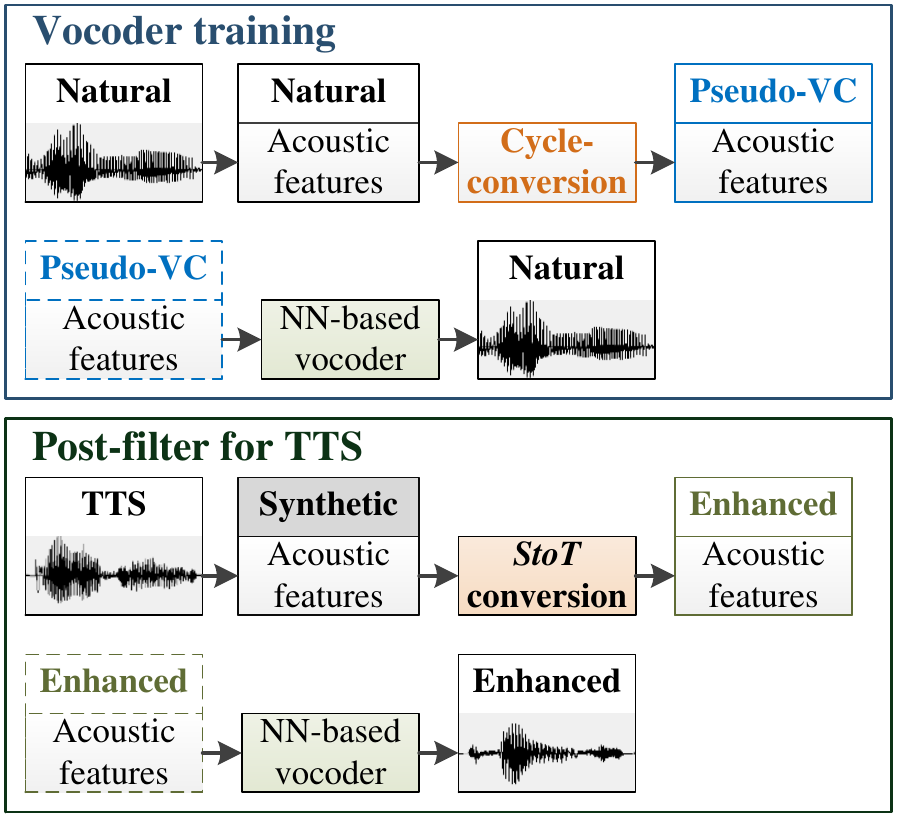}
\caption{Vocoder training and post-filter testing stages}
\label{fig:cycvc_pf}
\end{center}
\end{figure}

As shown in Fig.~\ref{fig:cycvc_train}, the synthetic and natural features are respectively taken as the source and target of a Cycle-VC model. The {\it StoT} module is trained using the synthetic-natural feature pairs to enhance the input synthetic features. For the pseudo-VC path, the cascaded {\it TtoS} and {\it StoT} modules are trained to reconstruct the input natural features. As shown in Fig.~\ref{fig:cycvc_pf}, after finishing the training of the Cycle-VC model, the natural features for training an NPF are first converted by the Cycle-VC model to get the pseudo-VC features. Then, the NPF is trained using the pairs of the pseudo-VC features and corresponding natural waveforms. In the NPF testing stage, the synthetic features extracted from TTS-generated speech are converted by the {\it StoT} module to get the enhanced features. The final enhanced speech is generated using the trained NPF conditioned on the enhanced features.

\section{Experiment with oracle phoneme duration}
\label{sec:experiment1}

\subsection{Corpus and TTS system}
To develop low-cost TTS systems and the proposed post-filtering framework, an internal Japanese corpus was adopted. A female and a male speakers were included in the corpus. Each speaker had 800 training and 100 testing utterances, and the average length of utterances was around 4 seconds. The sampling rate of this corpus was 48~kHz.

The acoustic features for the TTS systems were 60-dimensional mel-cepstral feature ($mcep$), one-dimensional log-scaled fundamental frequency ($F_0$), five-dimensional aperiodicity ($ap$), and their delta and delta-delta terms. All features were extracted using a WORLD vocoder~\cite{world}, and the minimum description length was set to 1.0. 

Four speaker-dependent (SD) low-cost TTS systems were involved in the evaluations. 
Two of the TTS systems were HMM-based, and the other two systems were DNN-based. The HMM-based systems were built using the Hidden Semi-Markov Model (HSMM) training script of HTS (ver. 2.3.2)~\cite{hts}. The phoneme HSMMs were initialized using manual phoneme segmentation. Each DNN-based system consisted of four independent feed-forward DNNs for respectively predicting the $F_0$, $ap$, $mcep$, and durations. All systems adopted a maximum likelihood parameter generation (MLPG) mechanism~\cite{vc_mlpg}.

To generate the final speech waveforms, the WORLD vocoder was also utilized in these low-cost TTS systems. The synthetic speech of the training and testing sets in this chapter were generated using the manually refined phoneme segmentation to make their phoneme durations match that of the natural speech. Furthermore, a traditional spectral post-filter~\cite{hts, vc_mlpg, hmm_pf}, which was included in the HTS demo, with an enhanced coefficient 1.4 ($\beta=0.4$) was only applied to the DNN-based TTS system (DNN-PF).

\subsection{Cycle-VC model and neural vocoder}
Because most of the current neural vocoders support high-fidelity speech generation with sampling rates only under 24~kHz, all utterances including natural and synthetic ones were downsampled to 24~kHz. The input of the Cycle-VC models and the auxiliary features for the WN and PWG vocoders included 45-dimensional $mcep$, one-dimensional log-scaled $F_0$ and unvoiced/voiced ($U/V$) binary code, and three-dimensional coded $ap$. Note that the outputs of the Cycle-VC models were only the $mcep$ features.

The settings of the Cycle-VC model followed our previous work~\cite{cyc_vc}, and the training epoch was set to 15. Several SD WN and PWG vocoders were involved in the evaluations to show the effectiveness of the proposed framework for different speakers, TTS systems, and neural vocoders. The architecture and training processing of the WN and PWG vocoders followed our previous work~\cite{qpnet_2021, qppwg_2021} with 200,000 and 400,000 iterations, respectively.

\begin{figure}[t]
\fontsize{8pt}{9pt}
\selectfont
{%
\begin{tabularx}{1.0\columnwidth}{@{}XX@{}}
\toprule
(a) Natural & (b) Synthetic \\ 
  \includegraphics[width=0.49\columnwidth]{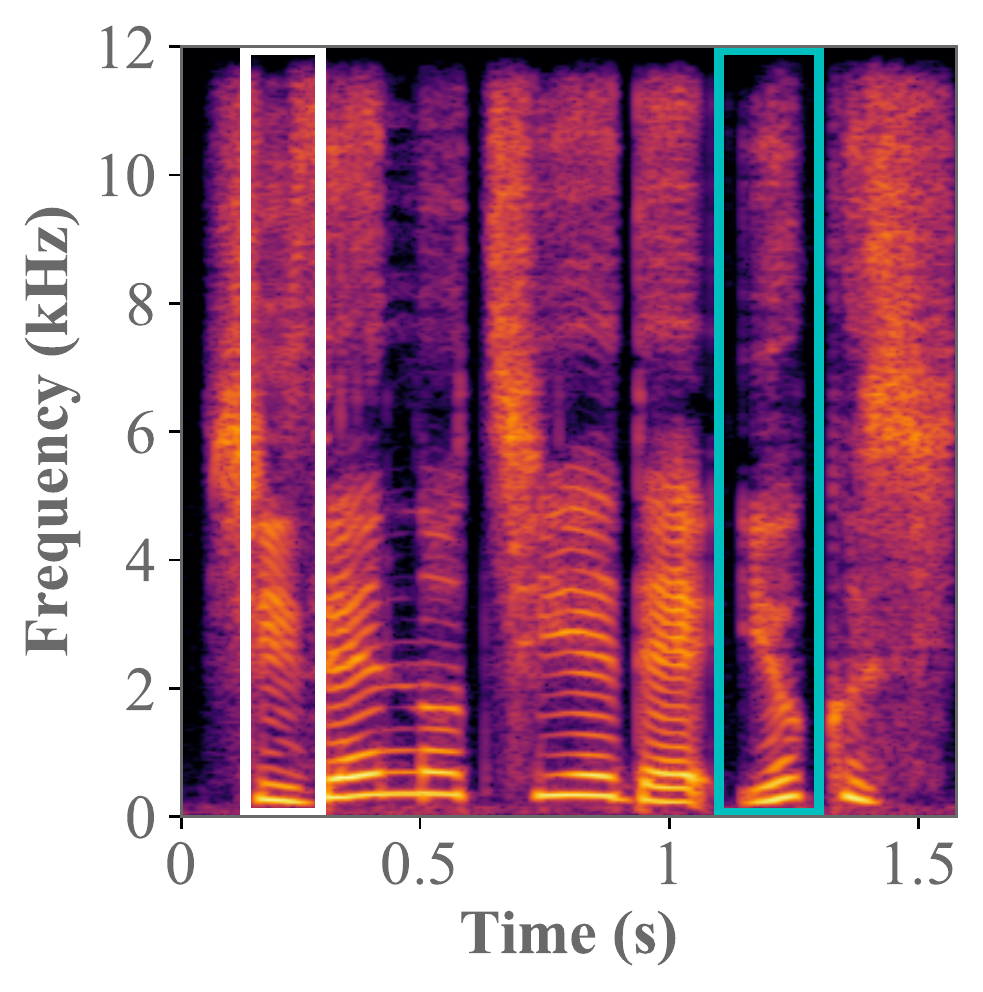}
&  \includegraphics[width=0.49\columnwidth]{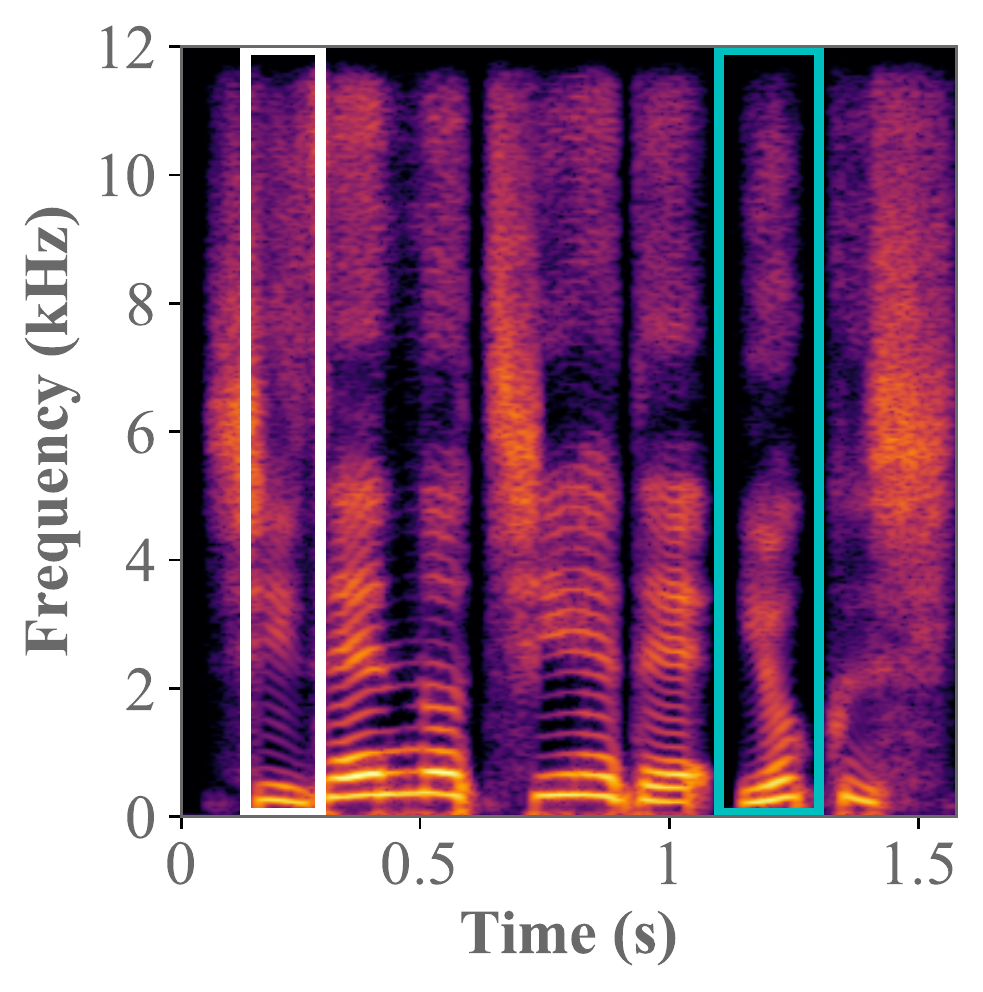} \\
(c) Pseudo-VC & (d) Enhanced \\ 
  \includegraphics[width=0.49\columnwidth]{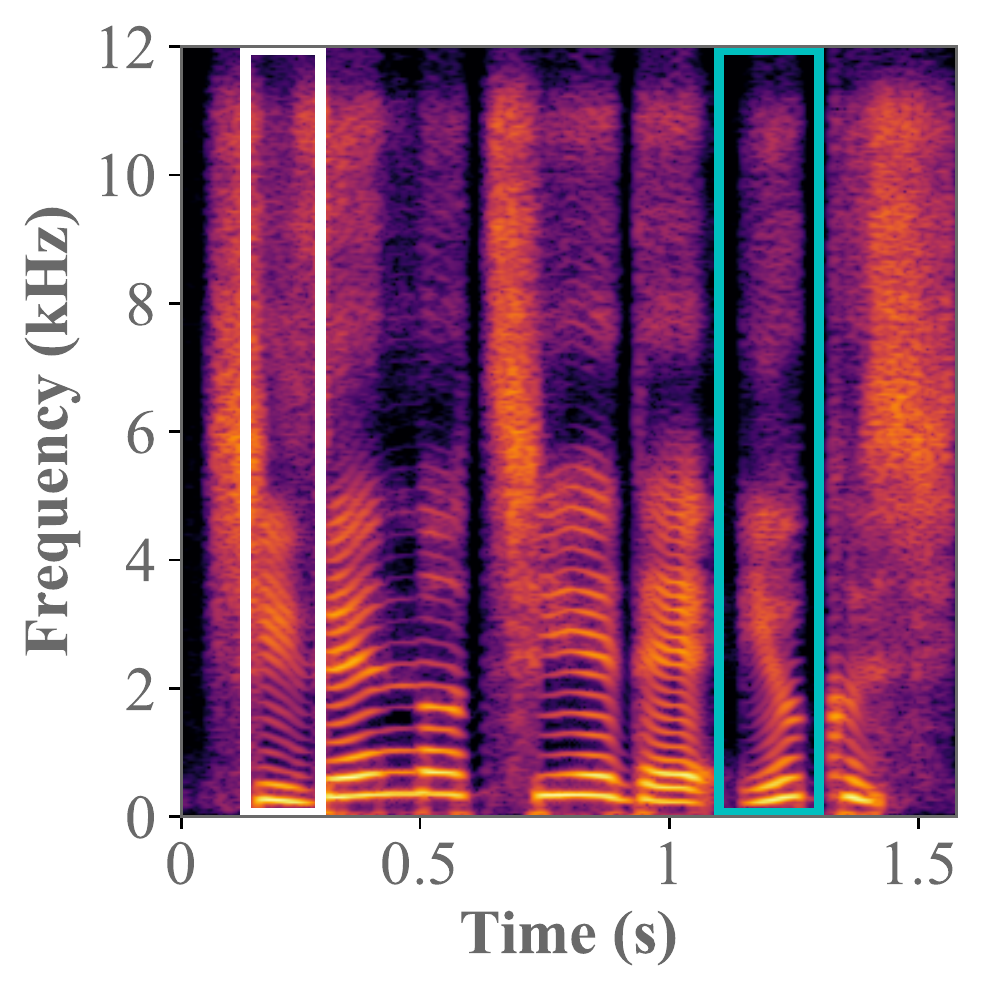}
& \includegraphics[width=0.49\columnwidth]{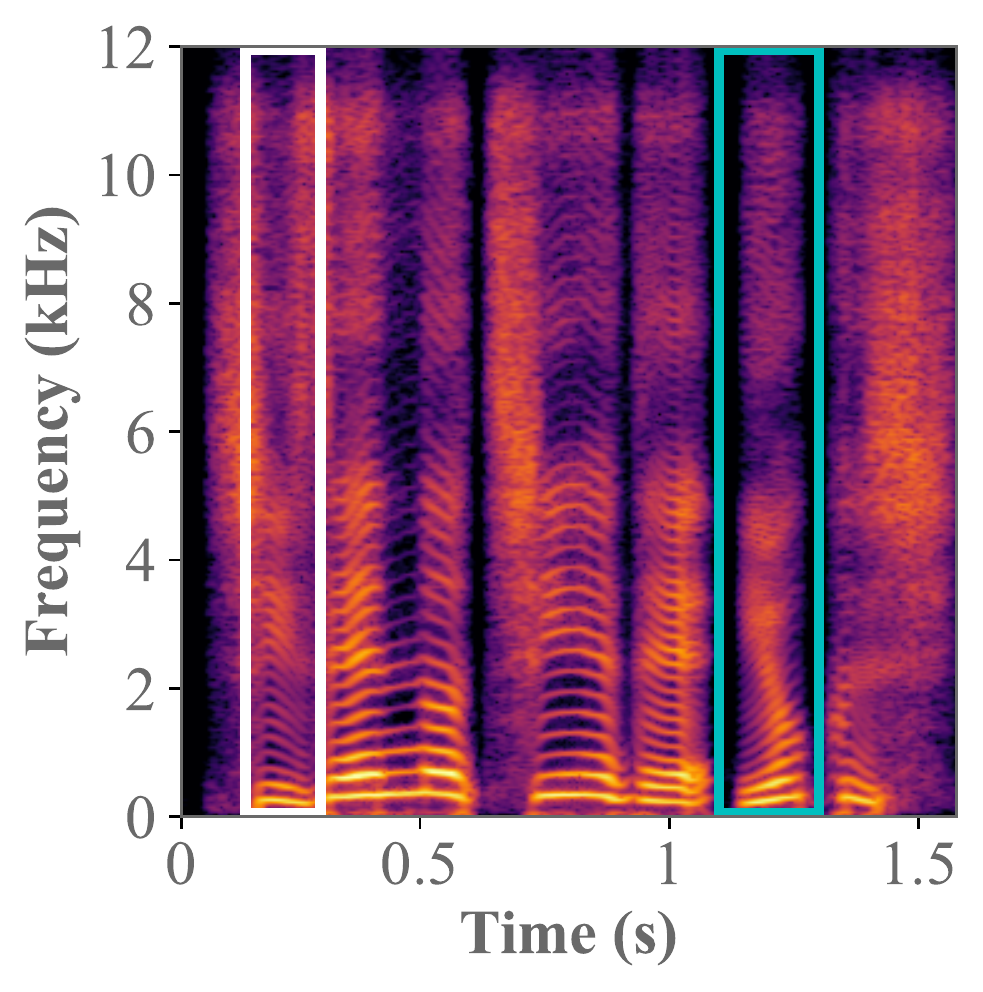} \\
\bottomrule
\end{tabularx}%
}
\caption{Natural, synthetic, pseudo-VC and enhanced spectrograms.}
\label{fig:spec}
\end{figure}

\subsection{Objective evaluation}
\label{sec:exp_obj}
In this section, the natural, synthetic, pseudo-VC, and enhanced spectrograms and power spectral densities (PSDs) are presented to visually show the TM and AM problems and the effectiveness of the proposed framework. After that, we objectively measure the similarities among natural, synthetic, and Cycle-VC processed spectral features using mel-cepstral distortion (MCD). Moreover, the log spectral distortions (LSD) and log GV distances (LGD) of the original TTS and post-filtering speech are also presented to show the effectiveness of the proposed mismatch refinements for the neural vocoders.

\subsubsection{Temporal and Acoustic Mismatches}

\begin{figure}[t]
\fontsize{8pt}{9pt}
\selectfont
{%
\begin{tabularx}{1.0\columnwidth}{@{}XX@{}}
\toprule
(a) Natural and Synthetic & (b) Pseudo-VC and Enhanced \\ 
  \includegraphics[width=0.49\columnwidth]{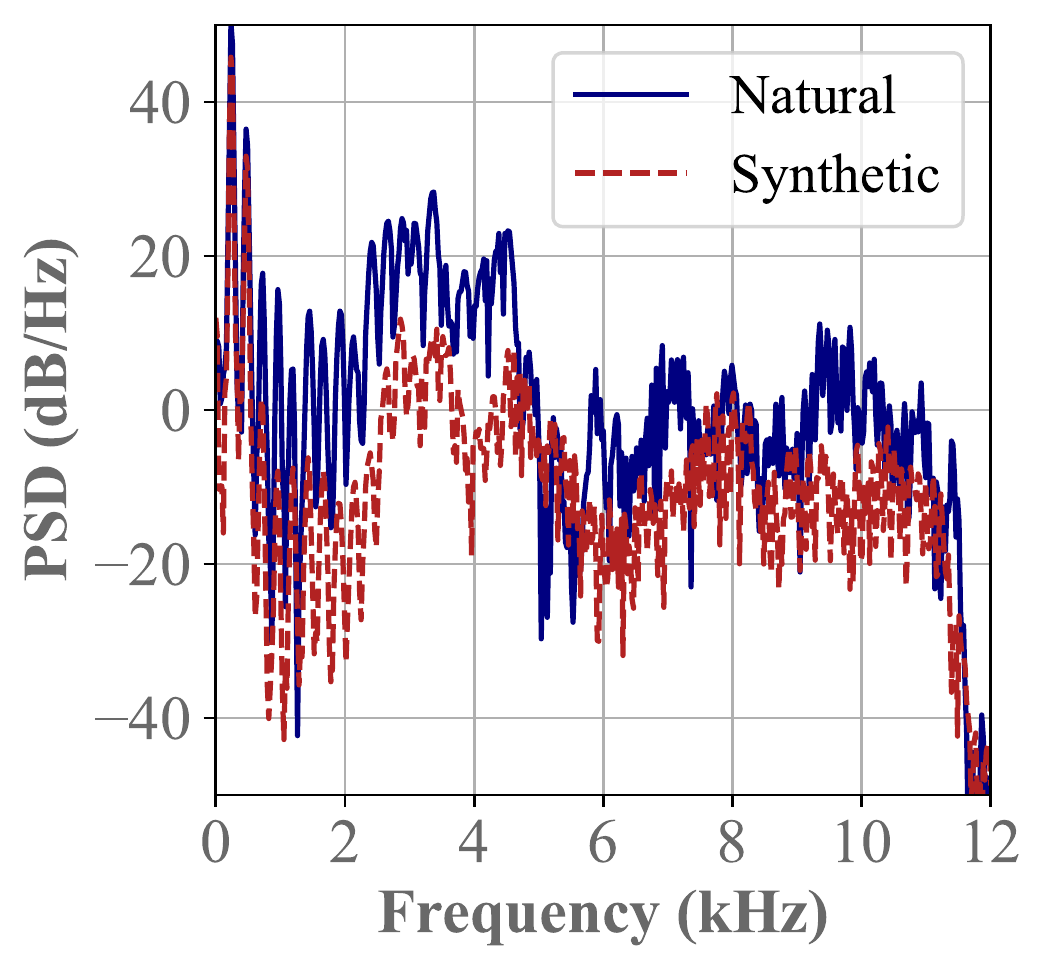}
&  \includegraphics[width=0.49\columnwidth]{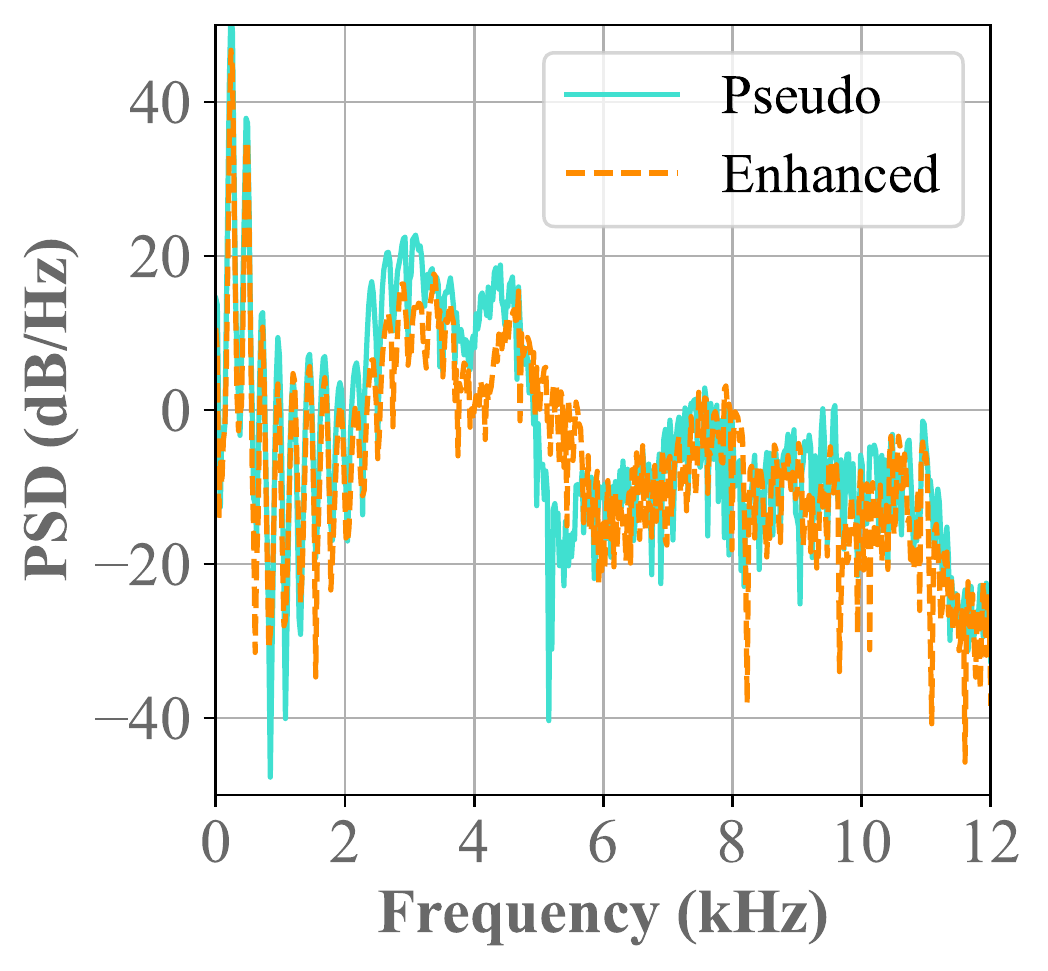} \\
\bottomrule
\end{tabularx}%
}
\caption{Comparison of the PSDs of Fig.~\ref{fig:spec} around 0.2~s (white boxes).}
\label{fig:psd}
\end{figure}

First, in this section, we argue that even if the oracle phoneme durations are given, the temporal structures of the synthetic and natural utterances are still different. Specifically, one phoneme is usually distributed to several continuous acoustic feature frames and more waveform samples. Since neural vocoders tackle speech modeling in the waveform domain, even very detailed temporal mismatches among a few waveform samples will cause performance degradations and error propagations, especially the AR vocoders. The performance degradations will be demonstrated in the following evaluation sections, and the visualized temporal mismatches are presented as follows. 

As shown in the white and cyan boxes of Fig.~\ref{fig:spec} (a) and (b), we can find that although the overall spectrograms of the natural and synthetic utterances are similar, there are many different details along the time axis. However, as shown in the boxes of Fig.~\ref{fig:spec} (c), the temporal structures of pseudo-VC spectrograms are more consistent with that of the natural utterance because the pseudo-VC spectral features are converted from the natural spectral features. Furthermore, as shown in Fig. (c) and (d), both the pseudo-VC and enhanced spectrograms suffer from the over-smoothing problem, so the characteristics of their spectrograms are similar. 

The PSDs around 0.2~s (the white boxes) of these four spectrograms shown in Fig~\ref{fig:spec} are also presented in Fig.~\ref{fig:psd}. We can find that the natural and synthetic PSDs are very different but the pseudo-VC and enhanced PSDs are similar. The pseudo-VC and enhanced PSDs are even more similar to the natural PSD than the synthetic PSD. To sum up, since the temporal structures of the natural and pseudo-VC utterances are coherent, training the neural vocoder using the pseudo-VC acoustic features and natural speech waveforms avoids the TM problem. Since the acoustic characteristics of the pseudo-VC and enhanced spectral features are similar, the training and testing mismatches (the AM problem) can be greatly eased.

\subsubsection{Comparison of Auxiliary Spectra}

In addition to the visualized results, to further objectively measure the similarities among the natural, synthetic, pseudo-VC, and enhanced spectrograms, we calculated the MCDs between each pair as shown in Table~\ref{tb:mcd}. All results are the average MCDs of the testing data respectively from the DNN- and HMM-based TTS systems, and the higher the MCD the lower the similarity of the spectrograms. The results of the DNN- and HMM-based systems have a very similar tendency. Therefore, in the following discussion, the DNN- and HMM-based prefixes are omitted, and their results are jointly discussed.

As shown in Table~\ref{tb:mcd}, the synthetic and natural spectrograms are the most different (with the highest MCDs), which indicates that training a vocoder using natural features but testing the vocoder using synthetic features will cause serious AM problems. The result also implies that the synthetic spectrograms have very different temporal structures from that of the natural spectrograms, and the TM problems obstruct training the vocoder using the synthetic features.

\begin{table}[t]
\caption{Mel-cepstral Distortion (dB) of Auxiliary Spectra}
\label{tb:mcd}
\fontsize{9pt}{10.8pt}
\selectfont
{%
\begin{tabularx}{\columnwidth}{@{}p{3.5cm}YY@{}}
\toprule
& DNN-based & HMM-based \\ \midrule
Synthetic to Natural     & 7.50 & 5.63 \\ \midrule
Enhanced to Pseudo-VC    & 3.75 & 4.11 \\
Pseudo-VC to Natural     & 3.30 & 4.26 \\ \midrule
Enhanced to Natural      & 5.01 & 5.21 \\ 
Enhanced to Synthetic    & 5.91 & 2.74 \\
Synthetic to Pseudo-VC   & 7.12 & 4.88 \\
\bottomrule
\end{tabularx}%
}
\end{table}

However, the pseudo-VC and enhanced spectrograms achieve much lower MCDs, which demonstrates the effectiveness of the proposed method to alleviate the AM problems. Furthermore, the pseudo-VC and natural spectrograms also achieve very low MCDs as expected, and the result confirms that the pseudo-VC features are more suitable for training a vocoder than the synthetic features because of the eased TM problems.

The results of other MCD pairs are also presented for reference. The low MCDs between the enhanced and natural spectrograms also show the effectiveness of the Cycle-VC module to enhance the synthetic spectral features. With the MCD results, the effectiveness of the proposed Cycle-VC module to alleviate the AM and TM problems of the auxiliary features for the final-stage neural vocoders is implied.

\subsubsection{Comparison of Final Spectra}
\begin{table}[t]
\caption{Comparison of Evaluated Vocoders}
\label{tb:vocoder}
\fontsize{9pt}{10.8pt}
\selectfont
{%
\begin{tabularx}{\columnwidth}{@{}p{1.5cm}YY@{}}
\toprule
        & \multicolumn{2}{c}{Acoustic features} \\ 
        & Training           & Testing \\ \midrule
UB      &                    & Natural   \\
DNN-AM  & Natural            & Synthetic \\
HMM-AM  &                    & Synthetic \\ \midrule
DNN-TM  & Synthetic          & Synthetic \\ 
HMM-TM  & Synthetic          & Synthetic \\ \midrule
DNN-NPF" & Pseudo   converted & Enhanced  \\ 
HMM-NPF & Pseudo   converted & Enhanced  \\ \bottomrule
\end{tabularx}%
}
\end{table}

To evaluate the overall performance of the proposed post-filter including the Cycle-VC and neural vocoder modules, the LSD and LGD of the unprocessed TTS synthetic and the post-filtered speech are presented. Specifically, there were two low-cost TTS systems, DNN- and HMM-based, as the lower bounds. Two different neural vocoders, WN and PWG, were adopted to generate the evaluation speech with seven conditions. As shown in Table~\ref{tb:vocoder}, the upper bound (UB) of the proposed framework is training and testing the neural vocoders using natural features. The neural vocoders being trained with natural features but tested with synthetic features is denoted as (TTS system)-AM, and the neural vocoders being trained and tested with synthetic features is denoted as (TTS system)-TM. The proposed method is denoted as (TTS system)-NPF. Since the proposed NPF was applied to the DNN-based system combined with the traditional spectral post-filter~\cite{hts, vc_mlpg, hmm_pf} (DNN-PF) in this chapter, we denote the DNN-based system with the cascaded traditional and neural-based post-filters as DNN-NPF". For the objective evaluations, the lower the LSD and LGD, the higher the spectral similarity to the natural spectra.

As shown in Table~\ref{tb:lsdlgd}, compared to the baseline TTS systems, the proposed NPF framework achieves lower LSD and LGD in most cases, which demonstrates the effectiveness of the proposed post-filters to enhance the low-cost TTS synthetic speech. On the other hand, compared to the systems suffering the AM problem, the proposed post-filters also attain better performances with lower LSD and LGD, which implies the quality degradations caused by the AM problem and the effectiveness of the proposed method to tackle the AM problem. Moreover, the proposed framework adopting WN as the vocoder also outperforms the systems suffering the TM problem. However, we find that the PWG vocoder is more robust to the TM problem, so the PWG-based post-filters with the TM problem achieve lower LSD and LGD than the proposed ones in some conditions. The possible reason is that the discriminator and the Non-AR mechanism make the PWG vocoder more robust to the TM problem. Since the proposed method still causes some aliasings to the pseudo-VC and enhanced features, the distortions from the TM problem might be less for the PWG vocoders.

\begin{table}[t]
\caption{Log Spectral Distortion and Log GV Distance of Final Spectra}
\label{tb:lsdlgd}
\fontsize{9pt}{10.8pt}
\selectfont
{%
\begin{tabularx}{\columnwidth}{@{}p{2cm}YYYY@{}}
\toprule
        & \multicolumn{2}{c}{LSD} & \multicolumn{2}{c}{LGD} \\ \midrule
DNN-PF (TTS)  & \multicolumn{2}{c}{1.242} & \multicolumn{2}{c}{1.055} \\
HMM (TTS)     & \multicolumn{2}{c}{1.159} & \multicolumn{2}{c}{1.338} \\ \midrule
        & WN     & PWG    & WN     & PWG \\ \midrule
UB      & 0.878  & 0.841  & 0.669  & 0.484  \\ \midrule
DNN-AM  & 1.175  & 1.181  & 1.239  & 1.109  \\
DNN-TM  & 1.164  & \textbf{1.029}  & 1.707  & \textbf{0.703}  \\ 
DNN-NPF" & \textbf{1.106}  & 1.039  & \textbf{1.177}  & 0.989  \\ \midrule
HMM-AM  & 1.119  & 1.129  & 1.647  & 1.250  \\ 
HMM-TM  & 1.389  & 1.140  & 1.917  & \textbf{0.882}  \\ 
HMM-NPF & \textbf{1.116}  & \textbf{1.102}  & \textbf{1.172}  & 1.199  \\ \bottomrule
\end{tabularx}%
}
\end{table}

\subsection{Subjective evaluation}
\label{sec:exp_sub}
Since objective measurements usually cannot directly reflect the real perceptual quality, subjective evaluations are essential for speech generation tasks. To subjectively evaluate the effectiveness of the proposed method, we first compared the proposed post-filters using WN vocoders with the unprocessed low-cost TTS systems. Then, we compared the proposed method using WN and PWG vocoders to show the versatility of the proposed framework. Moreover, we conducted preference tests to show the effectiveness of the proposed method for alleviating the AM and TM problems.

For each testing system and scenario, we randomly selected 50 testing utterances to form the subjective sets. Therefore, the set sizes of section~\ref{sec:exp_sub1},~\ref{sec:exp_sub2}, and~\ref{sec:exp_sub3} are 500, 600, and 600 utterances, respectively. Twenty subjects were involved in the evaluations, and half of them were native speakers. All subjects used the same device to do the tests in quiet environments. Each subject joined in one or two evaluations of the total three ones and evaluated a part of the corresponding subjective set. Each utterance in these subjective sets was at least evaluated by one subject. The final results were the average scores of the testing speakers.

In sections~\ref{sec:exp_sub1},~\ref{sec:exp_sub2}, we conducted mean opinion score (MOS) tests to evaluate the naturalness of the generated speech. Each subject was asked to give a score from 1 (worst) to 5 (best) to evaluate the speech quality. In section~\ref{sec:exp_sub3}, we conducted AB preference tests to compare the naturalness of each pair, which includes two generated speech utterances. Each subject was asked to listen to two utterances and then select the one with higher speech quality. Demo samples can be found on our demo page~\cite{demo}.

\subsubsection{Neural Post-filter for Low-cost TTS}
\label{sec:exp_sub1}
\begin{table}[t]
\caption{Naturalness in MOS (1--5) of Generated Speech w/ and w/o NPF}
\label{tb:mos_npf}
\fontsize{9pt}{10.8pt}
\selectfont
{%
\begin{tabularx}{\columnwidth}{@{}YY>{\centering}p{1.5cm}Y>{\centering\arraybackslash}p{1.5cm}@{}}
\toprule
UB & DNN-PF & DNN-NPF" & HMM & HMM-NPF \\ 
(WN)   &        & (WN)    &     & (WN)    \\ \midrule
4.49$\pm$.12 &2.90$\pm$.19 & 4.08$\pm$.15 &2.41$\pm$.16 & 3.71$\pm$.18 \\ \bottomrule
\end{tabularx}%
}
\end{table}
In this section, two low-cost TTS systems, DNN- and HMM-based, were considered. The DNN-based TTS system was combined with the additional traditional spectral post-filter~\cite{hts, vc_mlpg, hmm_pf}. As shown in Table~\ref{tb:mos_npf}, the proposed post-filters significantly enhanced the speech quality of the low-cost TTS-generated utterances. Specifically, even though the quality of the HMM-based TTS is quite low, the proposed post-filter markedly improve it with more than 1 in MOS. Although the traditional post-filter was applied to the DNN-based systems, which outperform the HMM-based systems, the proposed post-filter still markedly further improves their speech quality. Although there is still a quality gap between the UB and the post-filtering speech, the results confirm the effectiveness of the proposed post-filter for different low-cost TTS systems.

\subsubsection{AR and Non-AR Neural Post-filter}
\label{sec:exp_sub2}
\begin{table}[t]
\caption{Naturalness in MOS (1--5) of WN- and PWG-generated Speech}
\label{tb:mos_ar}
\fontsize{9pt}{10.8pt}
\selectfont
{%
\begin{tabularx}{\columnwidth}{@{}p{0.8cm}YYY@{}}
\toprule
    & UB           & DNN-NPF"      & HMM-NPF \\ \midrule
WN  & 3.94$\pm$.20 & 3.55$\pm$.18 & 3.10$\pm$.19 \\
PWG & 4.04$\pm$.20 & 3.52$\pm$.18 & 3.66$\pm$.19 \\ \bottomrule
\end{tabularx}%
}
\end{table}
After we confirmed the significant quality improvements of the proposed post-filter for different low-cost TTS systems, we explored the possibility of adopting different vocoders for the proposed framework. Specifically, the speech modeling capability of the AR-based WN and non-AR-based PWG was first evaluated. As shown in Table~\ref{tb:mos_ar}, we can find that both vocoders achieve comparable speech quality when the inputs are natural acoustic features. Then, we evaluated the performance of the proposed method respectively adopting the WN and PWG vocoders. According to the results in Table~\ref{tb:mos_ar}, we can find that the PWG vocoder is also suitable for the proposed method because of the comparable post-filtering quality for the DNN-based TTS and even higher post-filtering quality for the HMM-based TTS while compared to the post-filters with the WN vocoder. 

Furthermore, since only the DNN-based TTS adopted the conventional spectral post-filter in advance, the HMM-based TTS suffered the more severe oversmoothing problem. Although the proposed cyclical approach can ease the oversmoothing problem, the enhanced acoustic features still suffered the oversmoothing problem to some degree. However, we can find that the PWG-based NPF is more robust to the severe oversmmothing problem, and the result might be due to the effectiveness of the GAN mechanism to tackle the oversmoothing problem. More discussions can be found in the following section~\ref{sec:exp1_discussion}. In conclusion, the evaluation results demonstrate the proposed NPF is general for different vocoders.

\subsubsection{Mismatch Refinement}
After the effectiveness of the proposed post-filter adopting the WN and PWG vocoders are confirmed, we evaluated the effectiveness of the proposed framework to tackle the AM and TM problems. Specifically, we respectively compared the proposed post-filters (NPFs) with the post-filters suffering the AM and TM problems. As shown in Table~\ref{tb:pk_refine}, the WN-based post-filters are very vulnerable to the TM problems, and the possible reason might be the sample-by-sample modeling by the AR mechanism. Moreover, the AM problem also causes significant quality degradations to the WN-based post-filters since the proposed WN-based NPFs attain higher preferences while compared to the WN-based post-filters suffering the AM and TM problems.

On the other hand, the results of the PWG-based post-filters show that the TM problem also causes significant degradations even for the non-AR vocoder, and the proposed NPFs markedly outperforms the post-filters suffering the TM problem. However, the results in Table~\ref{tb:pk_refine} also show that the speech qualities of the NPFs and the post-filters suffering from the AM problem are comparable because their 95~\% confidence intervals overlap more than 50~\%. The result implies that the PWG-based post-filters are robust to the AM problem. Since the tendency of the subjective results is a little different from that of the objective results, we further listen to these generated samples and provide more insights in the following section.

\subsection{Discussion}
\label{sec:exp1_discussion}
After carefully listening to the generated samples, we find that stability and brightness are the two factors most related to perceptual quality. Specifically, unstable sound means that the generated speech suffers discontinuities such as wobbling sounds. Brightness refers to the upper mid and high-frequency speech contents. According to our observations, the TM problem is most related to stability, and the AM problem is most related to brightness. Moreover, we also find that humans are more sensitive to unstable sounds than over-smoothed sounds.

\label{sec:exp_sub3}
\begin{table}[t]
\caption{Naturalness Preference Evaluation Results (\%)}
\label{tb:pk_refine}
\fontsize{9pt}{10.8pt}
\selectfont
{%
\begin{tabularx}{\columnwidth}{@{}p{0.8cm}YYYY@{}}
\toprule
& DNN-NPF" / -AM & DNN-NPF" / -TM & HMM-NPF / -AM & HMM-NPF / -TM \\ \midrule
WN  
& \textbf{62}~/~38~$\pm$12 & \textbf{91}~/~9~$\pm$8
& \textbf{57}~/~43~$\pm$15 & \textbf{94}~/~6~$\pm$10 \\
PWG 
& 48~/~\textbf{52}~$\pm$13 & \textbf{60}~/~40~$\pm$12
& \textbf{54}~/~46~$\pm$14 & \textbf{85}~/~15~$\pm$12 \\ \bottomrule
\end{tabularx}%
}
\end{table}

For the NPFs with the WN vocoders, because of the AR manner, the temporal mismatched synthetic features and natural samples obstruct the WN vocoder to learn the correct sequential relationships among speech samples. Therefore, the WN-generated speech suffering the TM problem is very unstable that includes many broken segments, incorrect pronunciations, and unsteady loudness. The instability can be implied from both the objective and subjective evaluation results. Moreover, the over-smoothed synthetic features and the AM problem also result in quality degradations reflecting on the evaluation results.

For the NPFs with the PWG vocoders, because of the non-AR manner and the guides from the discriminator, the PWG-generated speech with the TM problem is more stable than that of the WN-generated ones. Since the PWG-generated speech with the TM problem is relatively stable and AM-free, while the proposed NPF-generated speech still suffers some acoustic mismatches, the PWG-generated speech with the TM problem achieves slightly lower LSDs and LGDs. However, because humans are very sensitive to speech instability, the PWG-generated speech with the TM problem still achieves lower preferences in the subjective evaluations. On the other hand, we find that although the PWG vocoders trained with natural acoustic features suffer the AM problem, the GAN structure eases the speech degradations. The result is consistent with the conclusion of previous literature~\cite{vawgan} that GAN is very effective to address the oversmoothing problem. Therefore, the quality improvements of the proposed NPF to tackle the AM problem become insignificant especially when the input TTS speech is already processed by a simple post-filter.

To sum up, both the TM and AM problems cause significant quality degradations when the neural post-filter adopts a WN vocoder, so the proposed framework achieves the significant objective and subjective improvements in WN-based NPFs. On the other hand, although the GAN structure and non-AR mechanism ease the TM and AM problems in PWG-based NPFs, the proposed framework still achieves comparable or better performance. Please refer to our demo page~\cite{demo} for more details.

\section{Experiment with predicted phoneme duration}
\label{sec:experiment2}
In the previous chapter, the effectiveness of the proposed framework using different neural vocoders for the synthetic speech with oracle phoneme duration has been demonstrated. In this chapter, we further explore the proposed framework in the real-world scenario that the phoneme duration of the synthetic speech in the testing stage is predicted.

The strategy of whether using the synthetic speech with the ground-truth phoneme duration to train the Cycle-VC model is first explored. Then, because the more challenging scenario causes some mismatches, the effectiveness of introducing external data for improving the robustness of the neural vocoder is explored. Furthermore, since we aim to build a practical NPF for the real-world scenario, the robustness of the proposed NPF for the TTS systems with different speech qualities is also evaluated in the following sections. 

\subsection{Experimental Settings}
The two internal Japanese target speakers with the same settings were adopted in the following evaluations, too. The differences are that the 800 utterances training dataset was divided into 750 utterances for training and 50 utterances for validation.
Moreover, four additional Japanese speakers’ data (two female and two male) from the same internal Japanese corpus and the VCTK~\cite{vctk} corpus were adopted as external data for pre-training the neural vocoders of the proposed NPFs. Each additional Japanese speaker also has 750 training and 50 validation utterances. The sampling rate of the VCTK corpus was converted to 24~kHz, and the number of the selected VCTK speakers was 96.

Since we focus on the effectiveness of the proposed framework in a real-world scenario, and the different TTS systems and vocoders have been evaluated in the previous chapter, only the DNN-based TTS systems and WN vocoders were adopted in the following evaluations. The architectures and hyperparameters of the DNN and WN models also followed that in the previous chapter. The only difference is that the WN vocoders were pre-trained using the natural acoustic features-waveforms pairs, which were the same as the WN vocoders with the AM problem, as their initial models. Then, the WN vocoders were adapted using the temporally matched pseudo-VC acoustic features and natural waveforms.

The architecture and hyperparameters of the Cycle-VC models in this chapter also followed the settings in the previous chapter. The only difference is that there were two versions of the synthetic utterances for training the Cycle-VC models because the ground-truth phoneme duration is still available in the training stage even for a real-world scenario. That is, in the training stage, the first version of the Cycle-VC models (CVC-m) adopted the synthetic utterances with the manually set phoneme duration. The second version of the Cycle-VC models (CVC-p) adopted the synthetic utterances with the predicted phoneme duration while the dynamic time warping (DTW) mechanism was applied for the acoustic feature alignments.

Preference tests were conducted to subjectively evaluate the naturalness of the generated speech for different training strategies. The 100 testing utterances of each speaker were divided into five non-overlapped sets, so each set included 20 female and 20 male testing utterances for each system. The number of subjects was 10, and they are native Japanese speakers. Each set was evaluated by two subjects. The final results were the averages of all sets, speakers, and subjects. 

Furthermore, MOS tests were conducted to evaluate the naturalness of the generated speech for different amounts of training utterances. We randomly selected 50 testing utterances from the 100 testing utterances for each speaker and divided the selected utterances into five non-overlapped sets. The number of native Japanese speaker subjects was also 10, and each set was also evaluated by two subjects. The final scores were the averages of all sets and subjects.

\subsection{Training strategies of the Cycle-VC and neural vocoding models}
A general real-world NPF framework includes training and testing stages. In the training stage, since we can prepare the paired synthetic and natural utterances, the ground-truth phoneme duration is available. However, in the testing stage, only the synthetic utterances are available, the phoneme duration should be predicted by the TTS system. As a result, there are two possible approaches to train a Cycle-VC model. 

Specifically, a Cycle-VC model can be trained using the synthetic speech with the manually set phoneme duration (CVC-m), but it will cause mismatches since the phoneme duration of the testing synthetic speech is predicted by the TTS model. The advantage of CVC-m is the stability of the training process. On the other hand, a Cycle-VC model can also be trained using the synthetic speech with the predicted phoneme duration (CVC-p), but the mismatches between the synthetic and natural utterances result in unstable Cycle-VC training. The instability of the Cycle-VC model will also degrade the stability of the downstream neural vocoder resulting in significant NPF quality degradations. However, if a CVC-p NPF can be successfully trained, the reduced training-testing mismatches of the Cycle-VC model might improve the NPF performance.

To improve the stability of the neural vocoder, increasing the amount and diversity of the training data is usually effective for the neural vocoders. Therefore, we also explored the effectiveness of pre-training the WN vocoder using the external corpus with many additional speakers. That is, the basic WN vocoders (w/o external) were pre-trained using the natural data of only the corresponding target speaker while the advanced WN vocoders (w/ external) were pre-trained using the external natural data of the four additional Japanese speakers and the selected 96 VCTK speakers.

As shown in Tables~\ref{tb:pk_paired} and~\ref{tb:pk_overall}, the naturalness of the generated utterances from the NPFs with pre-trained WNs, which suffer from the AM problems, CVC-m NPFs, and CVC-p NPFs were evaluated. The effectiveness of the large amount of external data for the WN pre-training was also evaluated. According to the results of the NPFs without using external pre-training data, we can find that the proposed CVC-m NPFs achieve the highest speech quality. However, because of the unstable issue, the speech quality of the CVC-p NPFs-generated speech significantly degrades. Moreover, the AM NPFs achieve similar speech qualities to the CVC-m NPFs, and the possible reason might be the training-testing mismatch of the Cycle-VC model training still degrades the performance of the CVC-m NPFs.

\begin{table}[t]
\caption{Paired Naturalness Preference Evaluation Results (\%) (-m: manual duration; -p: predicted duration for training)}
\label{tb:pk_paired}
\fontsize{9pt}{10.8pt}
\selectfont
{%
\begin{tabularx}{\columnwidth}{@{}p{1.7cm}YYY@{}}
\toprule
    & CVC-m / AM           & CVC-p / AM      & CVC-m / -p \\ \midrule
w/o external & \textbf{53}~/~47~$\pm$6 & 36~/~\textbf{64}~$\pm$9 & \textbf{68}~/~32~$\pm$7 \\
w/ external  & \textbf{67}~/~33~$\pm$8 & \textbf{68}~/~32~$\pm$9 & 45~/~\textbf{55}~$\pm$5 \\ \bottomrule
\end{tabularx}%
}
\end{table}

\begin{table}[t]
\caption{Overall Naturalness Preference Evaluation Results (\%) (-m: manual duration; -p: predicted duration for training)}
\label{tb:pk_overall}
\fontsize{9pt}{10.8pt}
\selectfont
{%
\begin{tabularx}{\columnwidth}{@{}p{1.7cm}YYY@{}}
\toprule
    & AM           & CVC-m      & CVC-p \\ \midrule
w/o external & 37.2$\pm$5.78 & \textbf{41.9}$\pm$4.53 & 20.9$\pm$5.62 \\
w/ external  & 19.8$\pm$5.93 & 36.7$\pm$4.00 & \textbf{43.5}$\pm$4.86 \\ \bottomrule
\end{tabularx}%
}
\end{table}

Furthermore, with the external pre-training corpus, both the CVC-m and CVC-p NPFs significantly outperform the NPFs suffering the AM problem, and the results demonstrate the effectiveness of the proposed framework to build an NPF for arbitrary low-cost TTS systems. The CVC-p NPFs even outperform the CVC-m NPFs showing that easing the training-testing mismatch of the Cycle-VC training improves the final performance. The possible explanation of the external pre-training corpus leading to the great improvements is that because of the higher capacity of the advanced pre-trained WN vocoder, the WN vocoder is capable to handle high diverse acoustic features and faithfully generates the corresponding speech according to the input acoustic features. Therefore, with the well pre-trained WN vocoder, the stability of the CVC-p NPFs has been improved, and the quality differences between CVC NPF- and AM NPF-generated utterances become larger.

\begin{table}[t]
\caption{Naturalness in MOS (1--5) of the Female Speaker Systems with Different Amounts of Training Utterances (50--750 utterances)}
\label{tb:mos_female}
\fontsize{9pt}{10.8pt}
\selectfont
{%
\begin{tabularx}{\columnwidth}{@{}p{1.5cm}YYYYY@{}}
\toprule
Natural & & 4.97$\pm$.06 & \\ \midrule
& T50 & T150 & T350 & T750 \\ \midrule
DNN & 2.34$\pm$.38 & 2.65$\pm$.39 & 2.96$\pm$.30 & 2.93$\pm$.29 \\ 
DNN-PF & 2.77$\pm$.34 & 3.20$\pm$.34 & 3.49$\pm$.33 & 3.49$\pm$.26 \\
DNN-NPF & \textbf{2.96}$\pm$.37 & \textbf{3.48}$\pm$.26 & \textbf{4.12}$\pm$.27 & \textbf{3.94}$\pm$.27 \\ \bottomrule
\end{tabularx}%
}
\end{table}

\subsection{Robustness to different TTS qualities}
To build practical real-world NPFs for arbitrary TTS systems, the robustness to the quality variations of different TTS systems is important. In this section, the CVC-p NPFs with the external pre-training corpus are adopted for several different TTS systems because of the best performance according to Tables~\ref{tb:pk_paired} and~\ref{tb:pk_overall}. That is, a multi-speaker WN vocoder was first built using the natural data of the four additional Japanese and 96 selected VCTK speakers. Then, the pre-trained WN vocoder was adapted for the different DNN-based TTS systems, which were trained using different amounts of training utterances from 50 to 750, to explore the robustness to TTS quality variations. Furthermore, the numbers of the generated pseudo-VC utterances for adaptation followed the same numbers of the TTS training utterances. For instance, if the TTS model was trained using 350 utterances, the CVC-p NPF was trained using 350 pseudo-VC utterances, too.

\begin{table}[t]
\caption{Naturalness in MOS (1--5) of the Male Speaker Systems with Different Amounts of Training Utterances (50--750 utterances)}
\label{tb:mos_male}
\fontsize{9pt}{10.8pt}
\selectfont
{%
\begin{tabularx}{\columnwidth}{@{}p{1.5cm}YYYYY@{}}
\toprule
Natural & & 4.92$\pm$.10 & \\ \midrule
& T50 & T150 & T350 & T750 \\ \midrule
DNN & 2.33$\pm$.46 & 2.65$\pm$.49 & 2.87$\pm$.38 & 3.10$\pm$.53 \\ 
DNN-PF & 2.78$\pm$.48 & 3.14$\pm$.57 & \textbf{3.56}$\pm$.50 & \textbf{3.82}$\pm$.37 \\
DNN-NPF & \textbf{2.87}$\pm$.49 & \textbf{3.15}$\pm$.48 & 3.38$\pm$.36 & 3.57$\pm$.25 \\ \bottomrule
\end{tabularx}%
}
\end{table}

Two baselines and natural utterances were taken as references. Specifically, both the DNN-based TTS systems without and with the traditional spectral post-filter~\cite{hts, vc_mlpg, hmm_pf} were the baselines. The proposed NPFs were applied to the DNN-based TTS systems without the traditional spectral post-filter because of the performance comparisons of the proposed NPF and traditional PF.

As shown in Tables~\ref{tb:mos_female} and~\ref{tb:mos_male}, the speech quality of the low-cost DNN-based TTS system is intuitively related to the amount of the training data. Although the speech qualities of the target female TTS systems saturate after the training data amount reaches 350 utterances. The overall tendency of the target female and male TTS systems is that the more the training utterances, the higher the MOS scores of the generated utterances. Moreover, after applying the proposed NPFs, the speech qualities of all TTS systems are significantly improved, which shows the effectiveness and robustness of the proposed NPFs for TTS quality variations.

On the other hand, compared with the traditional spectral PF, the proposed NPFs still outperform the traditional PF in most cases, which demonstrates the advanced speech enhancement capability of the proposed NPFs. Although the proposed NPFs achieve slightly lower MOS scores than that of the traditional PF for the male TTS systems trained with more than 350 utterances, the final speech quality might be still improved via applying the proposed NPFs to the DNN-PF systems since the cascaded PF-NPF (NPF") structure has sown the effectiveness in the previous chapter. To sum up, the proposed NPF is robust to arbitrary TTS systems with high speech quality variations.

\section{Conclusion}
\label{sec:conclusion}
In this paper, spectral enhancement and downstream vocoder modules are jointly considered in the proposed framework for the neural post-filtering of low-cost TTS systems. Specifically, we first argue that the conventional neural post-filtering frameworks, which independently train the spectral enhancement and vocoder modules, suffer two main problems, AM and TM, causing significant quality degradations. Then, to tackle the AM problem, we propose a Cycle-VC module to generate the pseudo-VC and enhanced features for respectively training and testing the proposed NPFs. Since both the pseudo-VC and enhanced features are processed by the Cycle-VC module, they have similar acoustic characteristics. Therefore, the training and testing mismatches can be eased. Furthermore, training the downstream vocoder using the pseudo-VC features instead of the synthetic features alleviates the TM between the input acoustic features and the output natural speech waveforms of the vocoder.

To show the proposed framework is general for arbitrary low-cost TTS systems and neural vocoders, DNN- and HMM-based TTS systems are adopted as the baseline TTS systems, and the AR-based WN and Non-AR-based PWG vocoders are used in the proposed NPFs. Both the objective and subjective evaluation results demonstrate that the TM and AM problems cause speech quality degradations for different TTS systems and vocoders, and the proposed framework alleviates the TM and AM problems to achieve higher speech quality. In conclusion, we explore the AM and TM problems for developing an NPF including spectral enhancement and vocoder modules and propose a general cyclical framework to tackle these problems for different TTS systems and neural vocoders.

Furthermore, to investigate the performance of the proposed NPF for the real-world scenario that the phoneme duration is predicted, we explored several training strategies for training the Cycle-VC and neural vocoder modules of the proposed NPF framework. We also explored the effectiveness and robustness of the proposed NPF for different TTS systems with speech quality and data quantity variations. In conclusion, with a well pre-trained neural vocoder, the proposed NPF still can work well for the real-world scenario even the quality of the TTS system is very low.

For future work, since there is a significant quality gap between the upper bound and the proposed method, we intend to investigate advanced frameworks to tackle the AM and TM problems such as different acoustic or self-supervised features or different network structures. Moreover, in addition to the CNN-based vocoders, we also intend to extend our work to more different vocoders such as WaveRNN and LPCNet.

\section*{Financial Support}

This work was partly supported by Japan Science and Technology Agency (JST), Core Research for Evolutional Science and Technology (CREST) (grant number JPMJCR19A3); and Japan Society for the Promotion of Science (JSPS) Grants-in-Aid for Scientific Research (KAKENHI) (grant numbers JP21H05054).

\section*{Statement of interest}
None.

\bibliography{mybib}
\bibliographystyle{IEEEtran}

\vskip2pc

\noindent \large \textbf{Biographies} 

\vskip2pc

\noindent\normalsize\textbf{Yi-Chiao Wu} received the B.S. and M.S. degrees in engineering from the School of Communication Engineering, National Chiao Tung University, Hsinchu, Taiwan, in 2009 and 2011, respectively, and the Ph.D. degree from the Graduate School of Informatics, Nagoya University, Nagoya, Japan, in 2021. He was with Realtek, ASUS, and Academia Sinica for five years. He is currently working as a Researcher. His research focuses on speech generation applications based on machine learning methods, such as neural vocoding, voice conversion, and speech enhancement.

\vskip2pc

\noindent\textbf{Patrick Lumban Tobing} (Member, IEEE) received the B.E. degree from the Bandung Institute of Technology, Bandung, Indonesia, in 2014, the M.E. degree from the Nara Institute of Science and Technology, Ikoma, Japan, in 2016, and the Ph.D. degree from the Graduate School of Information Science, Nagoya University, Nagoya, Japan, in 2020. He is currently working as a Researcher. He was the recipient of the Best Student Presentation Award from the Acoustical Society of Japan (ASJ). He is a Member of the IEEE and ISCA. 

\vskip2pc

\noindent\textbf{Kazuki Yasuhara} received the bachelor degree of engineering and the master degree of informatics from Nagoya University, Japan, in 2019 and 2021, respectively.
He is currently a reseacher at AI Inc.
His research interests include statistical approaches to speech and audio processing. 

\vskip2pc

\noindent\textbf{Noriyuki Matsunaga} received the B.E., M.E. and Ph.D. degrees in engineering from Toyama Prefectural University in 2009, 2011 and 2021, respectively. He is currently a researcher at research and development department at the AI, Inc. He is a member of the Acoustical Society of Japan, and the Institute of Electronics, Information and Communication Engineers.

\vskip2pc

\noindent\textbf{Yamato Ohtani} (Member, IEEE) received the B.E. degree in engineering from Osaka University, Japan, in 2005 and the M.E. and D.E. degrees from the Nara Institute of Science and Technology (NAIST) in 2007 and 2010, respectively. He was an intern researcher at ATR Spoken Language Communication Research Labratory, Kyoto, Japan from 2006 to 2009. He was a researcher at Toshiba Corporation 2010 to 2017. He was a senior researcher from 2017 to 2019 and he is currently the head of R\&D department at AI Inc. Since 2021, he has also been a director of AI Inc. His research interests include statistical approaches to speech and audio processing such as speech synthesis and voice conversion. He received the 39th Awaya Prize Young Researcher Award from the Acoustical Society of Japan (ASJ) in 2015.

\vskip2pc

\noindent\textbf{Tomoki Toda} (Senior Member, IEEE) received the B.E. degree from Nagoya University, Japan, in 1999, and the M.E. and D.E. degrees from the Nara Institute of Science and Technology (NAIST), Japan, in 2001 and 2003, respectively. He was an Assistant Professor from 2005 to 2011 and an Associate Professor from 2011 to 2015 with NAIST. Since 2015, he has been a Professor with the Information Technology Center, Nagoya University. His research interests include statistical approaches to speech and audio processing. From 2003 to 2005, he was a Research Fellow of the Japan Society for the Promotion of Science. He received more than ten article/achievement awards, including the IEEE SPS 2009 Young Author Best Paper Award and the 2013 EURASIP-ISCA Best Paper Award (Speech Communication journal).

\vskip2pc

\end{document}